\documentclass[iop]{emulateapj}

\def\na{New A}                
\def\actaa{Acta Astron.}      

\def\chandra{\emph{Chandra}}
\def\hubble{\emph{Hubble}}
\def\xmm{\emph{XMM-Newton}}
\def\rosat{\emph{ROSAT}}

\shorttitle{A dwarf nova in \object{M13}}
\shortauthors{Servillat et al.}

\begin{document}


\title{A dwarf nova in the globular cluster \object{M13}}




\author{M. Servillat\altaffilmark{1}, N. A. Webb\altaffilmark{2,3}, F. Lewis\altaffilmark{4,5}, C. Knigge\altaffilmark{6}, \\M. van den Berg\altaffilmark{7,1}, A. Dieball\altaffilmark{6} and J. Grindlay\altaffilmark{1}}
\affil{\altaffilmark{1}Harvard-Smithsonian Center for Astrophysics, 60 Garden Street, MS-67, Cambridge, MA 02138 \\{mservillat@cfa.harvard.edu}}
\affil{\altaffilmark{2}Universit\'e de Toulouse; Université Paul Sabatier - Observatoire Midi-Pyr\'en\'ees, \\Institut de Recherche en Astrophysique et Planétologie (IRAP), Toulouse, France}
\affil{\altaffilmark{3}Centre National de la Recherche Scientiﬁque; IRAP; 9 Avenue du Colonel Roche, BP 44346, F-31028 Toulouse cedex 4, France}
\affil{\altaffilmark{4}Department of Physics and Astronomy, The Open University, Walton Hall, Milton Keynes, MK7 6AA, UK}
\affil{\altaffilmark{5}Faulkes Telescope Project, University of Glamorgan, Pontypridd, CF37 4BD}
\affil{\altaffilmark{6}Department of Physics, University of Southampton, Southampton, S017 1BJ, UK}
\affil{\altaffilmark{7}Department of Physics and Astronomy, Utrecht University, 3508 TC Utrecht, The Netherlands}






\begin{abstract}
Dwarf novae in globular clusters seem to be rare with only 13 detections in the 157 known Galactic globular clusters.
We report the identification of a new dwarf nova in \object{M13}, the 14th dwarf nova identified in a globular cluster to date.
Using the 2m Faulkes Telescope North, we conducted a search for stars in \object{M13} that show variability over a year (2005--2006) on timescales of days and months.
This led to the detection of one dwarf nova showing several outbursts.
A \chandra\ X-ray source is coincident with this dwarf nova and shows both a spectrum and variability consistent with that expected from a dwarf nova, thus supporting the identification. 
We searched for a counterpart in \hubble\ Space Telescope ACS/WFC archived images and found at least 11 candidates, of which we could characterize only the 7 brightest, including one with a 3$\sigma$ H$\alpha$ excess and a faint blue star. 
The detection of one dwarf nova when more could have been expected likely indicates that our knowledge of the global Galactic population of cataclysmic variables is too limited. The proportion of dwarf novae may be lower than found in catalogs, or they may have a much smaller mean duty cycle ($\sim$1\%) as proposed by some population synthesis models and recent observations in the field.

\end{abstract}


\keywords{Galaxy: globular clusters: individual: \mbox{M13 (NGC 6205)} --
             X-rays: general --
             Stars: dwarf novae --
             Stars: novae, cataclysmic variables}




\section{Introduction}

Globular clusters (GCs) are old, gravitationally bound stellar systems with extremely high stellar densities in their core which enable cluster members to dynamically interact through regular encounters.
Close binary stars have been suggested to play a significant role in the dynamical evolution of GCs \citep{Hut+92}. Encounters between passing stars and close binaries may harden the latter and accelerate the passing star. This leads to the formation of a variety of tight and interacting binaries, and by accelerating stars in the core, close binaries may delay or halt the inevitable gravitational core collapse of GCs.

Many of the X-ray sources in GCs discovered in the 70's have been resolved into a multitude of close binaries and related systems with \xmm\ and \chandra\ \citep[see e.g.][]{SWB08,Servillat+08b}. These include active and quiescent low-mass X-ray binaries (LMXBs) with a neutron star (NS) primary, millisecond pulsars (MSPs), cataclysmic variables (CVs) and magnetically active binaries.
The number of NS LMXBs in GCs as well as part of the CV population appears to scale with the stellar encounter rate of their host cluster, indicating that they are indeed produced through dynamical mechanisms \citep{GBW03,Pooley+03,PH06}. The same correlation is observed with MSPs in GCs detected in radio \citep{hui10}, and even in $\gamma$-rays \citep{abdo10_msps_gcs}.

CVs are semi-detached binary stars with orbital periods typically of the order of hours, consisting of a white dwarf primary accreting via Roche lobe overflow from a companion which is usually a late-type, generally main-sequence star. Subtypes of CVs include non-magnetic CVs where an accretion disc forms around the primary, and magnetic systems where the white dwarf magnetic field is strong enough to disrupt (polar CVs, $B>10^7$~G) or truncate (intermediate polars, $10^5<B<10^7$~G) the accretion disc, and the mass flow is then channeled along the magnetic field lines \citep[e.g.][]{Cropper90,Patterson94}.
For systems with an accretion disc, it is believed that a thermal instability \citep[e.g.][]{Osaki96} is the cause of repetitive outbursts observed in CVs called dwarf novae (DNe).
Optically, DN eruptions have amplitudes of 2--6 magnitudes in V, a duration of a few to 20 days and a recurrence timescale of weeks to years \citep[e.g.][]{Osaki96}.

From the Catalog and Atlas of Cataclysmic Variables \citep[updated 2006]{Downes+01}, DNe are the most common CVs identified in the field, with about half of the identified CVs being DNe. However, the sample of known field CVs suffers from various selection effects \citep{PKK07}, and is biased towards bright and highly variable systems, such as DNe. We note that the distribution of those CVs is almost isotropic, which would indicate that the known population is mainly local (less than 1 kpc around the Sun). 
The global Galactic population and its intrinsic composition is thus not fully understood.

Population synthesis models indicate that a fainter population showing fewer outbursts may dominate the population of DNe \citep[up to 70\%,][]{kolb93,howell+97,willems05}. Recent detections in the SDSS of faint CVs with short periods showed one of the first hints for the existence of such a population \citep{gaensicke09}, but its proportion is still not constrained.

If we limit our knowledge to the locally known population of CVs, there appear to be fewer DN outbursts from CVs in GCs \citep{Shara+96,DLM06}. 
This would mean that the population of CVs in GCs is intrinsically different from the known population in the field. 
However, it is clear that the known population in the field is not a reliable reference. 
We note that \citet{Shara+96} based their comparison on a small sample of about 20 well studied DNe which are not likely to represent the global population.
Two hypotheses remain to be fully tested. (i) Dynamical effects and a specific environment in GCs (low metallicity, long timescales) in GCs could shape differently the population of CVs. For example, it has been proposed that CVs in GCs are dominated by moderately magnetic systems (intermediate polars) with low accretion rates which show fewer outbursts \citep{DLM06,Grindlay99}. Also, typical masses of the accreting primaries may be higher, as massive white dwarfs can be more efficiently placed in binaries by tidal capture and 3-body processes \citep{Ivanova+06}. (ii) The lack of DN outbursts from CVs in GCs could be due to detection biases. Due to the crowding in optical, DN outbursts are indeed more difficult to catch in GCs \citep{Pietrukowicz+08}.
These questions can only be solved with further observations and comparisons with more complete populations of CVs, both in the field and in GCs.

We present here a search for optically variable stars, and particularly DN outbursts, in the GC \object{M13}  (\object{NGC 6205}).
From the catalog of \citet[][updated in 2010]{Harris96}, \object{M13} is at a distance of 7.1~kpc from the Sun. Its core radius is $0\farcm62$, which corresponds to the half width half-maximum of the luminosity profile of the cluster, and its half-mass radius is $1\farcm69$ (region encircling half of the mass, approximated to be half of the light from the cluster). \object{M13} is a low metallicity cluster with $\mathrm{[Fe/H]}=-1.53$. The reddening along the line of sight is $\mathrm{E(B-V)} = 0.02$, which corresponds to a low extinction A$_V\sim0.06$ and an absorption from a column density of $\mathrm{N_H} \sim 1.1\times10^{20}$~cm$^2$.

We first present our search for optical variables and outbursts (\S\ref{obs}). We then detail \chandra\ X-ray observations of M13 used to search for X-ray sources coincident with the optically variable ones (\S\ref{chandra}). In \S\ref{DN}, we describe one peculiar optical event aligned with an X-ray source and present archived \hubble\ Space Telescope data to uncover possible counterparts. Finally, we discuss the results in \S\ref{discussion}.



\section{Optical Observations}
\label{obs}

\subsection{Data and reduction}

The GC \object{M13} has been observed with the Faulkes Telescope North (FTN) 
at Haleakala on Maui, USA, during the nights indicated in Table~\ref{tab:obs}. 
The FTN is a 2m robotic telescope optimized for research and education \citep{2008ca07.conf..504R,2009arXiv0902.4809L} and is part of the Las Cumbres Observatory Global Telescope Network.

Imaging was obtained in U, B, V, R, i, H$\alpha$ and O\texttt{III} filters between February 2005 and April 2006. We focus here on the U band as this dataset is the cleanest and the most complete, and CVs are generally blue objects so they are likely to be among the brightest objects compared to GC main sequence stars in the U band. Moreover, narrow band filters are not as deep, and we obtained fewer images in the other filters.

The DillCam CCD detector we used has $2048\times2048$ pixels binned $2\times2$ prior to
readout into effectively $1024\times1024$ pixels.
The field of view is $4\farcm7\times4\farcm7$ and the pixel scale is
$0\farcs278$ per pixel. Automatic pipelines de-bias and flatfield
the Faulkes science images using calibration files from
the beginning and the end of each night.

The seeing ranged from $1\farcs2$ to more than $3''$.
We discarded
images with poor signal-to-noise ratio (S/N; mostly due to
thin clouds) and images with tracking problems (where the
stars are non-spherical and stretch more than a few pixels
during the exposure), an excess of hot pixels, or a background
gradient across the image.
Finally, 78 U-band images were kept. They can be grouped into 5 epochs of observations, each epoch lasting one to a few days. The first 4 epochs were taken over two and a half months and the fifth epoch about one year later.

\begin{deluxetable}{cccl}
\tablecaption{Faulkes Telescope North U-band observations.\label{tab:obs}}
\tablewidth{0pt}
\tablehead{\colhead{Epoch} & \colhead{Date} & \colhead{MJD} & \colhead{Exposures}
}
\startdata
1 & 2005-02-18  & 53419 & 60s,120s \\
  & 2005-02-22  & 53423 & 120s \\
2 & 2005-03-02  & 53431 & 3x120s \\
  & 2005-03-04  & 53433 & 120s \\
3 & 2005-04-12  & 53472 & 8x120s \\
  & 2005-04-15  & 53475 & 9x120s \\
  & 2005-04-16  & 53476 & 9x120s \\
  & 2005-04-17  & 53477 & 9x120s \\
4 & 2005-05-05  & 53495 & 2x800s \\
5 & 2006-03-12  & 53806 & 44x200s
\enddata 
\end{deluxetable}

\subsection{Search for variability}

\begin{figure}[tbp]
\centering
\includegraphics[width=\columnwidth]{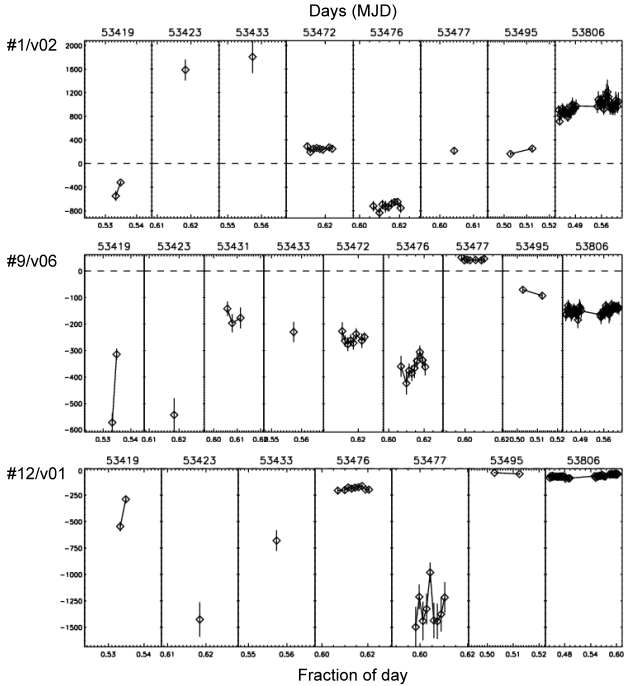}
\caption{Lightcurves of 3 variable star candidates associated with BL Her variables, see Table~\ref{tab:var}. The Y axis gives the flux relative to the reference image in arbitrary units with one sigma error bars.\label{fig:lc}}
\end{figure}

\begin{figure}[tbp]
\centering
\includegraphics[width=\columnwidth]{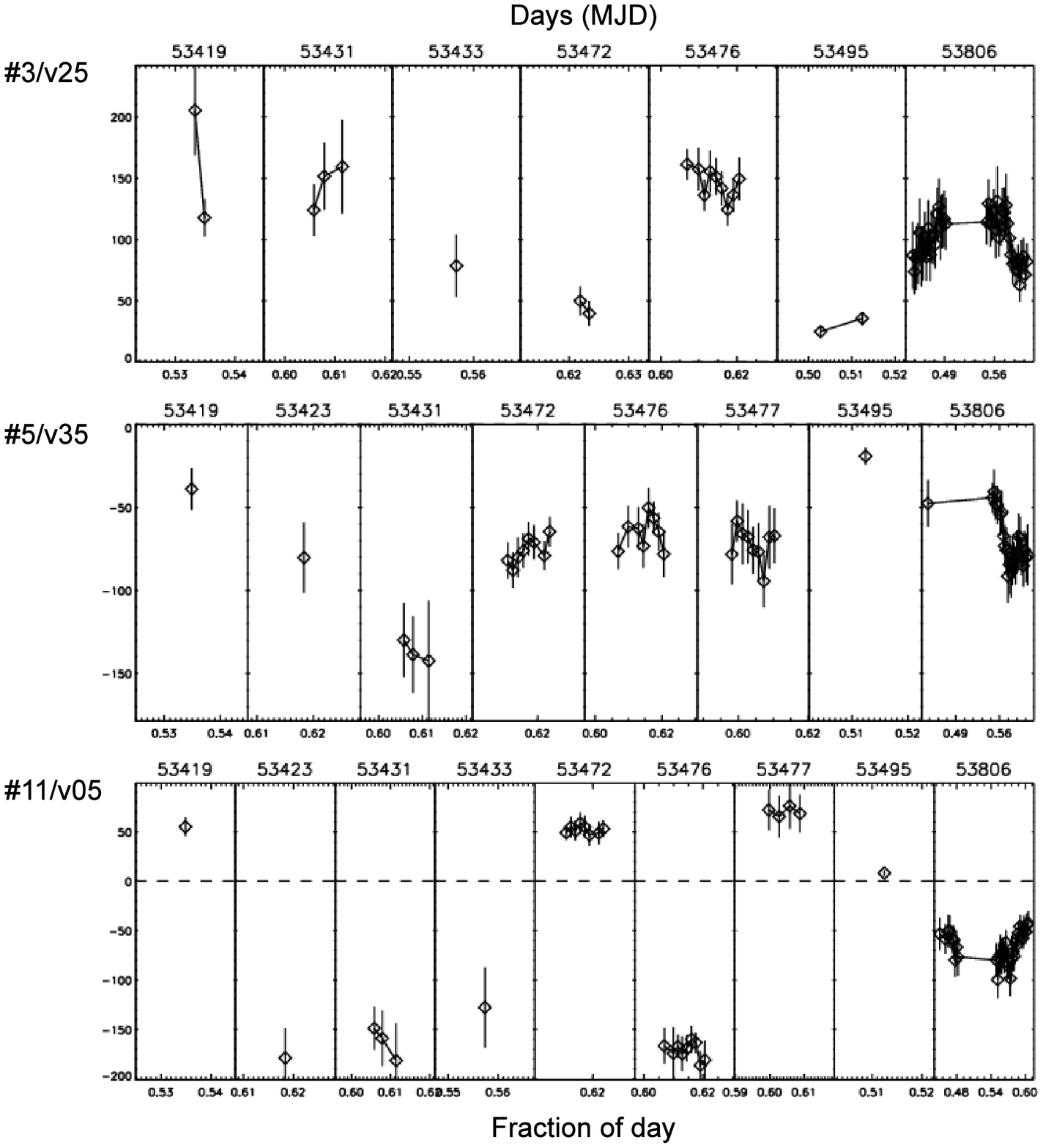}
\caption{Lightcurves of 3 variable star candidates associated with RR Lyr stars, same comments as in Figure~\ref{fig:lc}. \label{fig:lc1}}
\end{figure}

\begin{figure}[tbp]
\centering
\includegraphics[width=\columnwidth]{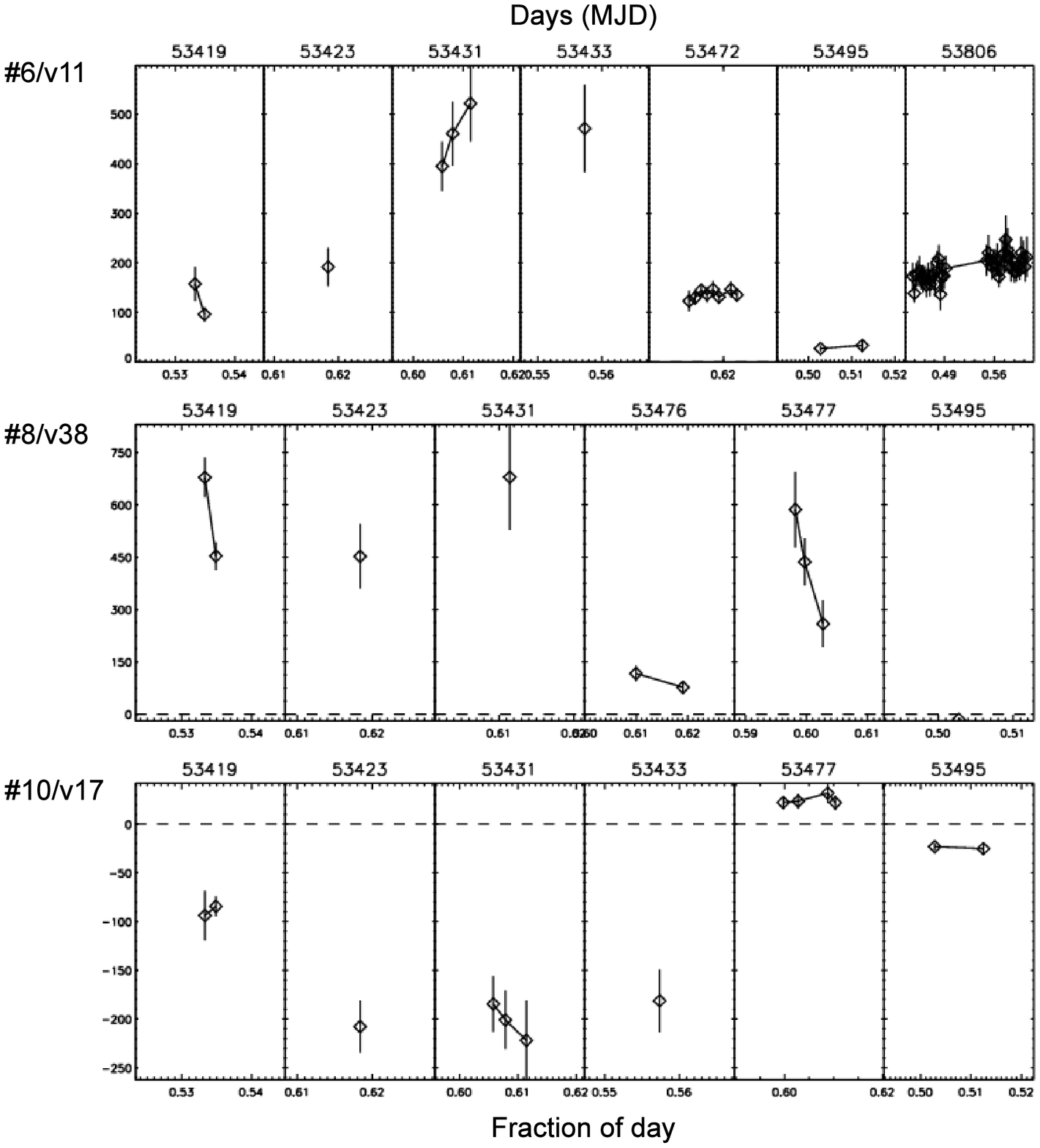}
\caption{Lightcurves of 3 variable star candidates associated with RG stars, same comments as in Figure~\ref{fig:lc}. \label{fig:lc2}}
\end{figure}

\begin{figure}[tbp]
\centering
\includegraphics[width=\columnwidth]{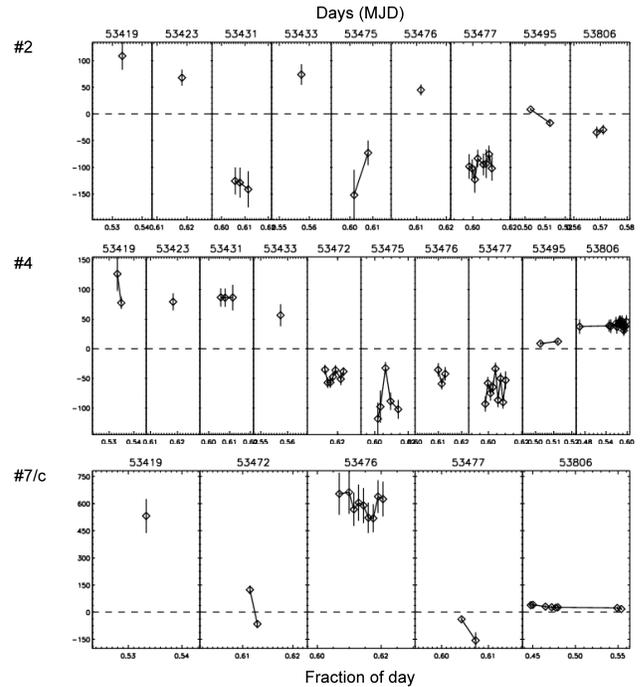}
\caption{Lightcurves of 3 variable star candidates not associated with previously known variable stars, see Figure~\ref{fig:lc}. Previously, star 7 had only been suspected to be variable. \label{fig:lc3}}
\end{figure}

We looked for variable stars in \object{M13} using difference imaging photometry. First, we aligned the images by selecting 3 stars in a reference image. The task \textit{imregister} under IRAF\footnote{IRAF is distributed by the National Optical Astronomy Observatory, which is operated by the Association of Universities for Research in Astronomy (AURA) under cooperative agreement with the National Science Foundation} was used to get a set of $1024\times1024$ pixels images all aligned on the reference image after rotation and translation. Due to this method, the edges of some images were cut, and the usable field of view for this search has been reduced to $\sim4'\times3\farcm7$, which still included the half-mass radius of \object{M13}.

We computed a reference image by taking the median of 5 images with the best seeing (about $1\farcs2$). This image was then subtracted from all the other images using the ISIS 2 software \citep{2000A&AS..144..363A}. This tool uses adaptive kernel techniques \citep{1998ApJ...503..325A} and gives optimal results in the case of crowded fields, variable seeing and point spread function (PSF).
We set the kernel size to $9\times9$ and the stamp size to $15\times15$ pixels (the stamp size is the area taken by the program around each object for computation). Three Gaussians were used in the convolution. The background is nearly flat so it was fitted with a constant. The spatial variations in the PSF kernel were modeled by a 4th order polynomial. After this process, the mean of the obtained images was close to zero, as expected, with some isolated positive or negative sources.

\begin{figure*}[tbp]
\centering
\includegraphics[width=.9\textwidth,trim = 30mm 15mm 40mm 8mm]{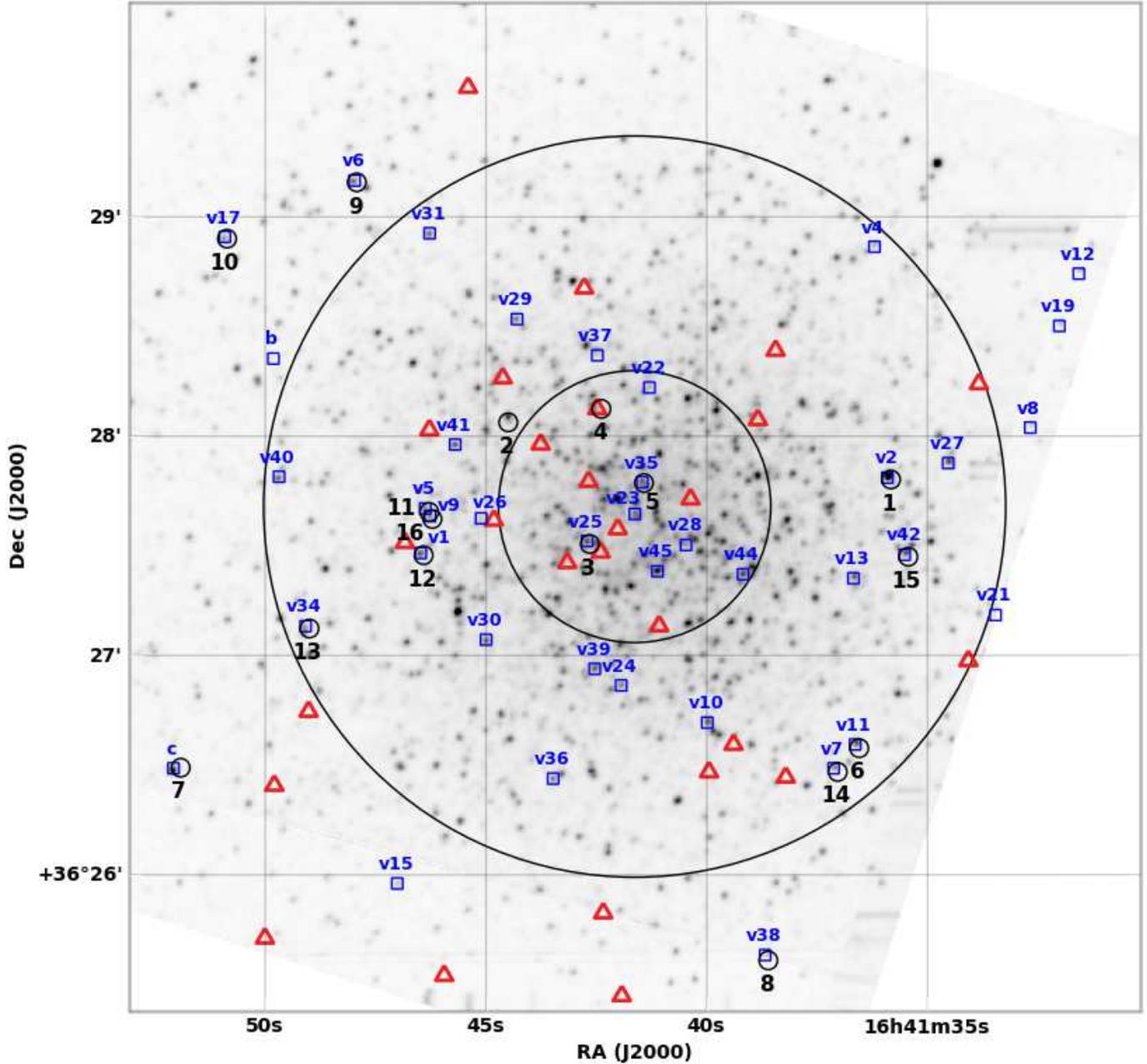}
\caption{M13 observed with the FTN (reference image). The core and half-mass radii are shown in black. Known variable objects from \citet{2003A&A...398..541K} are plotted as blue squares, X-ray sources detected with \chandra\ as red triangles and variable stars detected in this work as black circles.}
\label{fig:map}
\end{figure*}

\begin{deluxetable*}{cccccccc}
\tablecaption{List of variable stars.\label{tab:var}}
\tablewidth{0pt}
\tablehead{\colhead{\#} & \colhead{RA} & \colhead{Dec} & \colhead{Name} & \colhead{Type} & \colhead{Period} & \colhead{$<$V$>$} & \colhead{$\Delta$V}\\
   & [h:m:s] & [d:m:s] &  &  & [day] &  &
}
\startdata
 1 & 16:41:35.90 & 36:27:48.7 & v02 & BL Her &  5.11 & 13.01 & 0.87 \\
 2 & 16:41:44.57 & 36:28:04.4 & & & & &  \\
 3 & 16:41:42.73 & 36:27:31.1 & v25 & RR Lyr &  0.43 &     -- & 0.44 \\
 4 & 16:41:42.46 & 36:28:08.1 & & X-ray & & &  \\
 5 & 16:41:41.50 & 36:27:47.8 & v35 & RR Lyr &  0.32 &     -- & 0.25 \\
 6 & 16:41:36.62 & 36:26:35.2 & v11 &     RG & 91.77 & 11.93 & 0.13 \\
 7 & 16:41:51.99 & 36:26:29.9 &   c &     U-V & & &  \\
 8 & 16:41:38.68 & 36:25:37.2 & v38 &     RG &    -- & 12.12 & 0.07 \\
 9 & 16:41:48.00 & 36:29:10.1 & v06 & BL Her &  2.11 & 14.08 & 0.60 \\
10 & 16:41:50.95 & 36:28:54.6 & v17 &     RG & 43.04 & 11.98 & 0.38 \\
11 & 16:41:46.36 & 36:27:39.9 & v05 & RR Lyr &  0.38 & 14.77 & 0.42 \\
12 & 16:41:46.49 & 36:27:28.1 & v01 & BL Her &  1.46 & 14.09 & 1.04 \\
13 & 16:41:49.07 & 36:27:08.0 & v34 & RR Lyr &  0.39 & 14.83 & 0.32 \\
14 & 16:41:37.11 & 36:26:28.7 & v07 & RR Lyr &  0.31 & 14.93 & 0.30 \\
15 & 16:41:35.51 & 36:27:27.6 & v42 &     RG &    -- & 11.94 & 0.10 \\
16 & 16:41:46.29 & 36:27:37.9 & v09 & RR Lyr &  0.32 & 14.82 & 0.43
\enddata
\tablecomments{Variable stars detected with their coordinates (columns 2 and 3). When available, we list the name in column 4, and properties of the stars derived by \citet{2003A&A...398..541K} in columns 5 to 8: the type of star, the period, the average Vega magnitude in the V band, and the amplitude of the variability $\Delta V$ observed. Star 7 is labelled ''c'' as first reported by \citet{Meinunger78} and reused by \citet{2003A&A...398..541K}. RG stands for red giant and U-V for U-V-bright object as studied by \citet{ZNG72}. X-ray indicates an association with an X-ray source. RR~Lyr are short-period (0.2 to 2 days) population II pulsating variables often found in GCs. BL~Her are a subclass of Type II Cepheid variables with periods between 1 and 8 days.}
\end{deluxetable*}

For each subtracted image, we detected variable star candidates using IDL procedures from the Astrolib\footnote{http://idlastro.gsfc.nasa.gov/} (e.g. \textit{find}). We estimated the background level and the standard deviation $\sigma$ in an annulus around the source and performed a detection for sources with a significance higher than 5$\sigma$. However, when irregular features were present in the subtracted images, the IDL/Astrolib \textit{find} procedure gave a high number of false detections. When this was the case, all the detections for that image were ignored. We thus selected the most secure detections and possibly missed low significance events. We finally merged the different lists and found a total of 16 variable star candidates in the field of view. They are listed in Table~\ref{tab:var}

For each detected variable star in each subtracted image, we first fitted the star with a Gaussian, and estimated the integrated flux in arbitrary units. The typical observed full width half-maximum (FWHM) value in the subtracted images is about $1\farcs5$, slightly larger than the FWHM of the reference image. We estimated the significance of the variable star candidates by comparing their amplitude with the  standard deviation of the surrounding background. For the lightcurve extraction, we kept only measurements that were higher than three times the standard deviation for each variable star.
Some examples of lightcurves are shown in Figures~\ref{fig:lc}, \ref{fig:lc1}, \ref{fig:lc2}, and \ref{fig:lc3}. The zero level in those plots corresponds to the flux of the star in the reference image. If no point is reported for a given day, it is either because the luminosity of the object is at the same level as in the reference image (then no emission is seen in the subtracted image), or the emission is not significant ($<3\sigma$) compared to the surrounding background of the subtracted image.
We thus have discarded some low significance measurements.

In order to check if our method could lead to false positives, we added manually 5 entries in the variable star candidate list (aligned with bright stars or randomly placed in the image). The chosen positions did not correspond to objects detected as variables. We applied the same extraction process for all 5 positions and we did not obtain any significant measurements, i.e. we did not obtain false detections.


\subsection{Known variable stars}

In order to compare our list of variable stars with previous catalogs, we added an astrometric solution to the reference image by aligning stars using the 2MASS Catalog \citep{Skrutskie+06} which provided 15 good reference stars. We used GAIA\footnote{GAIA was created by the now closed Starlink UK project and has been more recently supported by the Joint Astronomy Centre Hawaii funded again by the Particle Physics and Astronomy Research Council (PPARC) and more recently by its successor organisation the Science and Technology Facilities Council (STFC).} to fit the position of those stars and generate an astrometric solution for the reference image with an RMS error of $0\farcs2$. Taking into account the accuracy of 2MASS, the total astrometry over the field of view is $0\farcs3$ (1$\sigma$ error).

We compared our variable star list to the Catalog of Variable Stars in Globular Clusters \citep[][]{2001AJ....122.2587C,2003A&A...398..541K}.
We note that, as the subtracted images are convolved (and thus smoothed) during the subtraction process, the positional accuracy can be lower than for the reference image. We iteratively cross-matched the two lists to find a systematic offset. This bore-sight correction amounts to $0\farcs93$ and $0\farcs66$ in RA and Dec, respectively, with a 1$\sigma$ error of $0\farcs15$ between the 2 sets, which adds up to the astrometric error.
This led to the identification of 14 variable star candidates as already known variables. The name, type, period and magnitude of the known variable stars are reported in Table~\ref{tab:var}.
The 16 variable stars detected in this study as well as the known variables stars and the X-ray sources in the field of view are shown in Figure~\ref{fig:map}.

Although  our data are insufficient to determine a period, the variability observed for stars 1 (v02) and 9 (v06) presented in Figure~\ref{fig:lc} is consistent with the previously observed periods \citep[][]{2003A&A...398..541K}. The same trend is seen for star 3 (v25) in Figure~\ref{fig:lc} with a shorter period of 0.3 day and all the other lightcurves extracted.

Red giants (RG) stars have previously been observed with semi- or irregular variability \citep{2003A&A...398..541K}. The lightcurves we obtained for this category of candidates are consistent with this observation (Figure~\ref{fig:lc2}) but due to our limited sampling they could as well be periodic.

The variable star 7 is a known U and V bright object \citep[misleadingly called UV-bright by][]{ZNG72}, associated with \object{M13} based on radial velocity measurements.
The U and V bright stars referred by \citet{ZNG72} are in fact related to post-horizontal branch stars when plotted in a CMD (see their Figure~2).
Star 7 corresponds to L993 and IV-52 in the catalogs of \citet{Ludendorff05} and \citet{Arp55}, respectively and has a spectral type A5III-IV from SIMBAD\footnote{http://simbad.u-strasbg.fr/}. This star has been reported as possibly variable \citep{Meinunger78} but needed confirmation. No variability was reported by \citet[][star labelled c]{2003A&A...398..541K}, but we show its variability in this study, which seems to occur on timescales of a few days.

Finally, the stars~2 and 4 have no previous detections as variable stars in the optical. Those stars are unlikely to be periodic variable stars with periods 0.2 to 8 days as previous, more complete searches for such periodic stars did not detect them \citep[e.g.][]{2003A&A...398..541K}. However, we note that the star 2 is variable over short timescales of one or two days, which seems quite short compared to typical DN outbursts.
In order to constrain the nature of those new events, we searched for possible X-ray counterparts, as reported in the next section.

\section{\chandra\ X-ray observation of M13}
\label{chandra}

\begin{figure}[tbp]
\centering
\includegraphics[width=\columnwidth]{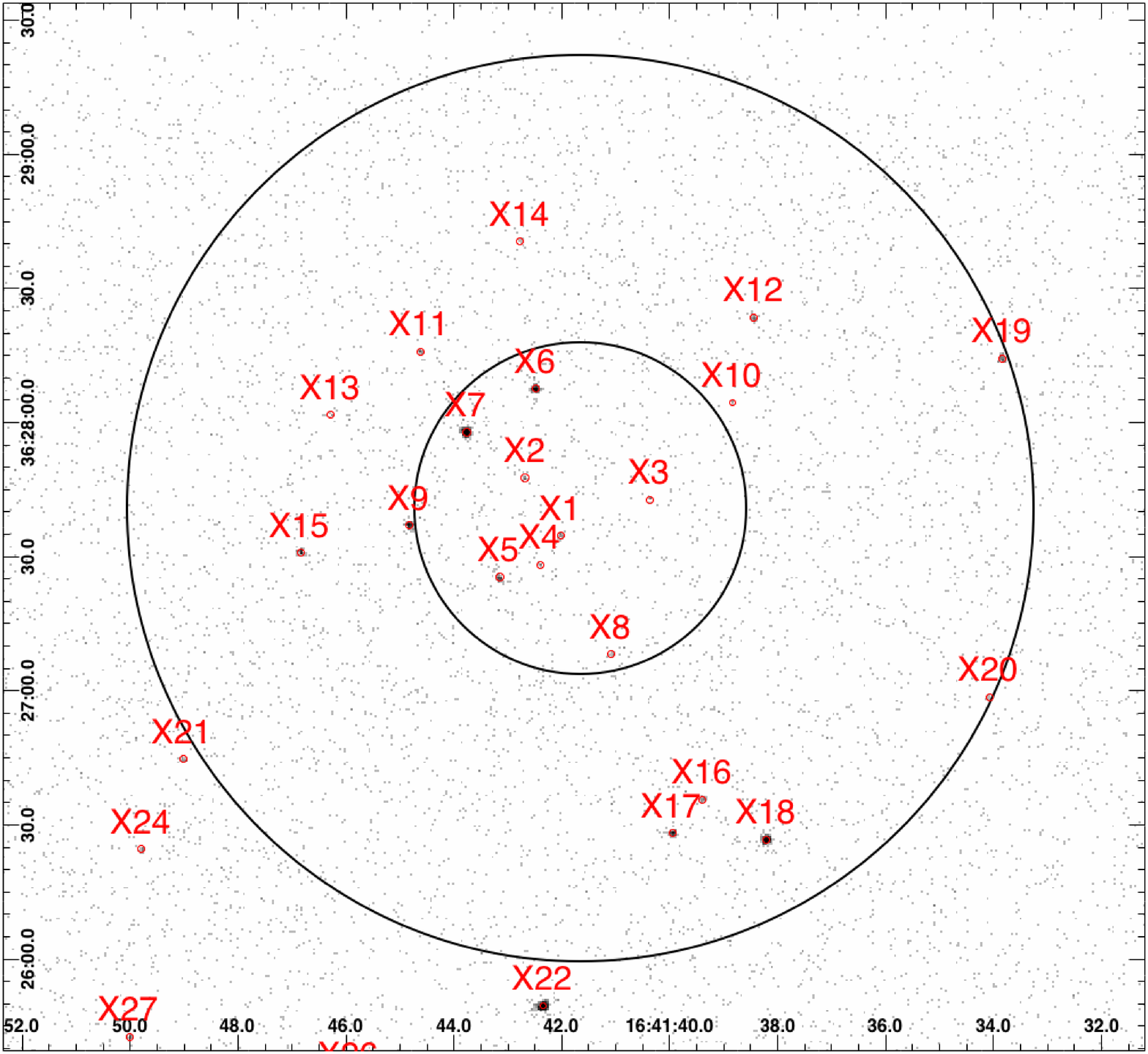}\\
\caption{\chandra\ image of M13 with circles corresponding to the core ($0\farcm62$) and half-mass radii ($1\farcm69$). Sources are labelled according to their distance to the center of the cluster. Labelled circles correspond to the 3$\sigma$ errors including the absolute pointing accuracy.}
\label{fig:xmap}
\end{figure}

\begin{figure}[tbp]
\centering
\includegraphics[width=\columnwidth]{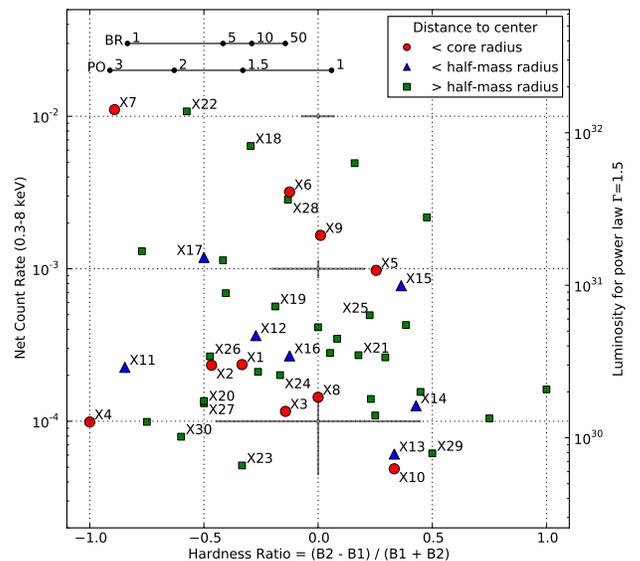}\\
\caption{Count rate vs. hardness ratio for the \chandra\ sources detected in the M13 field of view. For clarity, only the 30 X-ray sources closest to the center are labelled. Hardness ratios are computed from the counts in the bands $\mathrm{B1} = 0.3-1.5$~keV and $\mathrm{B2} = 1.5-8$~keV. Black solid lines indicate the hardness ratios corresponding to a power law model (PO) with different photon indices and a bremsstrahlung emission (BR) with different temperatures kT in keV. The luminosity for a source in the cluster with a spectrum modeled by a power law of photon index 1.5 is indicated on the right axis. The mean errors for 3 luminosity levels are indicated by gray bars.}
\label{fig:xhrd}
\end{figure}

\begin{deluxetable}{lccc}
\tablecaption{Detected X-ray sources and expected background sources in M13.\label{tab:xexpect} }
\tablewidth{0pt}
\tablehead{\colhead{Radius} & \colhead{$0\farcm62$} & \colhead{$1\farcm69$} & \colhead{$4\farcm2$}
}
\startdata
Detected & 8  & 15 & 35 \\
Expected & $0.3\pm0.3$  & $2.5\pm1.6$ & $15.4\pm3.9$ \\
[0.5ex] \tableline \\[-2ex]
Difference & $7.3\pm0.3$ & $12.5\pm1.6$ & $19.6\pm3.9$
\enddata 
\end{deluxetable}

\begin{deluxetable*}{ccccccccc}
\tabletypesize{\scriptsize}
\tablecaption{List of \chandra\ X-ray sources in the field of M13\label{tab:chsrclist}}
\tablewidth{0pt}
\tablehead{\colhead{Source ID} & \colhead{RA} & \colhead{Dec} & \colhead{95\% error} & \colhead{Distance} & \colhead{Net Count Rate} & \colhead{B1 Counts} & \colhead{B2 Counts} & \colhead{HR} \\
   & [h:m:s] & [d:m:s] & [''] & ['] & [$\times10^{-3}$ cnt s$^{-1}$] & 0.3--1.5 keV & 1.5--8 keV &
}
\startdata
X1  & 16:41:42.015 & 36:27:34.65 &   0.70 &   0.13 & 0.24$\pm$0.07 & 10 & 5 &  -0.33$\pm$0.36 \\ 
X2  & 16:41:42.676 & 36:27:47.65 &   0.68 &   0.24 & 0.23$\pm$0.07 & 11 & 4 &  -0.47$\pm$0.35 \\ 
X3  & 16:41:40.366 & 36:27:42.84 &   0.68 &   0.26 & 0.12$\pm$0.05 & 4 & 3 &  -0.14$\pm$0.53 \\ 
X4  & 16:41:42.401 & 36:27:28.29 &   0.50 &   0.26 & 0.10$\pm$0.04 & 6 & 0 &  -1.00$\pm$0.41 \\ 
X5  & 16:41:43.160 & 36:27:25.44 &   0.42 &   0.40 & 0.98$\pm$0.14 & 22 & 37 &   0.25$\pm$0.18 \\ 
X6  & 16:41:42.480 & 36:28:07.52 &   0.15 &   0.48 & 3.19$\pm$0.24 & 99 & 77 &  -0.12$\pm$0.11 \\ 
X7  & 16:41:43.761 & 36:27:57.84 &   0.08 &   0.51 & 11.05$\pm$0.45 & 575 & 33 &  -0.89$\pm$0.05 \\ 
X8  & 16:41:41.078 & 36:27:08.11 &   0.69 &   0.56 & 0.14$\pm$0.05 & 6 & 6 &   0.00$\pm$0.41 \\ 
X9  & 16:41:44.822 & 36:27:37.04 &   0.26 &   0.64 & 1.66$\pm$0.18 & 46 & 47 &   0.01$\pm$0.15 \\ 
X10 & 16:41:38.839 & 36:28:04.53 &   0.40 &   0.69 & 0.05$\pm$0.03 & 1 & 2 &   0.33$\pm$0.80 \\ 
X11 & 16:41:44.622 & 36:28:15.97 &   0.39 &   0.84 & 0.23$\pm$0.07 & 12 & 1 &  -0.85$\pm$0.34 \\ 
X12 & 16:41:38.439 & 36:28:23.48 &   0.55 &   0.96 & 0.36$\pm$0.08 & 14 & 8 &  -0.27$\pm$0.30 \\ 
X13 & 16:41:46.283 & 36:28:01.74 &   0.69 &   1.00 & 0.06$\pm$0.04 & 2 & 4 &   0.33$\pm$0.57 \\ 
X14 & 16:41:42.775 & 36:28:40.55 &   1.36 &   1.02 & 0.13$\pm$0.05 & 2 & 5 &   0.43$\pm$0.52 \\ 
X15 & 16:41:46.838 & 36:27:30.93 &   0.31 &   1.06 & 0.77$\pm$0.12 & 14 & 30 &   0.36$\pm$0.21 \\ 
X16 & 16:41:39.393 & 36:26:35.67 &   0.58 &   1.17 & 0.27$\pm$0.07 & 9 & 7 &  -0.12$\pm$0.35 \\ 
X17 & 16:41:39.950 & 36:26:28.15 &   0.31 &   1.26 & 1.19$\pm$0.15 & 51 & 17 &  -0.50$\pm$0.17 \\ 
X18 & 16:41:38.209 & 36:26:26.71 &   0.12 &   1.41 & 6.39$\pm$0.34 & 228 & 124 &  -0.30$\pm$0.07 \\ 
X19 & 16:41:33.832 & 36:28:14.33 &   0.41 &   1.67 & 0.57$\pm$0.10 & 19 & 13 &  -0.19$\pm$0.25 \\ 
X20 & 16:41:34.074 & 36:26:58.57 &   0.84 &   1.67 & 0.13$\pm$0.05 & 6 & 2 &  -0.50$\pm$0.48 \\ 
X21 & 16:41:49.019 & 36:26:44.73 &   0.65 &   1.75 & 0.27$\pm$0.07 & 7 & 10 &   0.18$\pm$0.34 \\ 
X22 & 16:41:42.346 & 36:25:49.57 &   0.09 &   1.86 & 10.76$\pm$0.45 & 460 & 124 &  -0.58$\pm$0.06 \\ 
X23 & 16:41:45.407 & 36:29:35.47 &   0.52 &   2.06 & 0.05$\pm$0.03 & 2 & 1 &  -0.33$\pm$0.80 \\ 
X24 & 16:41:49.794 & 36:26:24.47 &   0.62 &   2.08 & 0.20$\pm$0.06 & 7 & 5 &  -0.17$\pm$0.41 \\ 
X25 & 16:41:41.926 & 36:25:26.92 &   0.62 &   2.23 & 0.50$\pm$0.10 & 12 & 19 &   0.23$\pm$0.25 \\ 
X26 & 16:41:45.947 & 36:25:32.48 &   0.68 &   2.31 & 0.27$\pm$0.07 & 14 & 5 &  -0.47$\pm$0.31 \\ 
X27 & 16:41:50.006 & 36:25:42.67 &   0.65 &   2.59 & 0.14$\pm$0.05 & 6 & 2 &  -0.50$\pm$0.48 \\ 
X28 & 16:41:37.792 & 36:25:05.22 &   0.20 &   2.70 & 2.84$\pm$0.23 & 90 & 69 &  -0.13$\pm$0.11 \\ 
X29 & 16:41:27.839 & 36:27:34.26 &   0.90 &   2.78 & 0.06$\pm$0.04 & 1 & 3 &   0.50$\pm$0.68 \\ 
X30 & 16:41:41.646 & 36:24:41.70 &   1.44 &   2.98 & 0.08$\pm$0.04 & 4 & 1 &  -0.60$\pm$0.60 \\ 
X31 & 16:41:50.285 & 36:25:13.85 &   0.25 &   3.00 & 1.30$\pm$0.16 & 62 & 8 &  -0.77$\pm$0.15 \\ 
X32 & 16:41:54.727 & 36:26:04.15 &   0.75 &   3.09 & 0.43$\pm$0.09 & 8 & 18 &   0.38$\pm$0.27 \\ 
X33 & 16:41:45.948 & 36:24:22.01 &   0.62 &   3.42 & 0.69$\pm$0.12 & 33 & 14 &  -0.40$\pm$0.20 \\ 
X34 & 16:41:48.727 & 36:30:48.17 &   0.75 &   3.43 & 0.41$\pm$0.09 & 13 & 13 &   0.00$\pm$0.28 \\ 
X35 & 16:41:32.606 & 36:30:37.58 &   0.37 &   3.46 & 2.17$\pm$0.20 & 33 & 93 &   0.48$\pm$0.12 \\ 
X36 & 16:41:36.849 & 36:31:15.09 &   1.08 &   3.70 & 0.28$\pm$0.08 & 9 & 10 &   0.05$\pm$0.32 \\ 
X37 & 16:41:39.239 & 36:23:43.84 &   0.62 &   3.98 & 1.14$\pm$0.15 & 51 & 21 &  -0.42$\pm$0.16 \\ 
X38 & 16:41:30.540 & 36:31:02.52 &   1.59 &   4.03 & 0.16$\pm$0.06 & 0 & 10 &   1.00$\pm$0.32 \\ 
X39 & 16:41:39.846 & 36:23:31.32 &   1.20 &   4.17 & 0.21$\pm$0.07 & 12 & 7 &  -0.26$\pm$0.32 \\ 
X40 & 16:41:21.629 & 36:29:01.59 &   1.35 &   4.24 & 0.10$\pm$0.05 & 1 & 7 &   0.75$\pm$0.46 \\ 
X41 & 16:42:02.732 & 36:27:21.40 &   1.09 &   4.25 & 0.14$\pm$0.05 & 5 & 8 &   0.23$\pm$0.39 \\ 
X42 & 16:41:52.258 & 36:23:23.67 &   1.13 &   4.79 & 0.11$\pm$0.05 & 3 & 5 &   0.25$\pm$0.50 \\ 
X43 & 16:41:57.158 & 36:31:24.07 &   1.57 &   4.86 & 0.26$\pm$0.08 & 6 & 11 &   0.29$\pm$0.34 \\ 
X44 & 16:41:32.745 & 36:23:08.08 &   1.86 &   4.88 & 0.16$\pm$0.06 & 8 & 21 &   0.45$\pm$0.26 \\ 
X45 & 16:42:03.911 & 36:25:27.95 &   1.60 &   5.00 & 0.10$\pm$0.04 & 7 & 1 &  -0.75$\pm$0.46 \\ 
X46 & 16:42:02.659 & 36:30:59.96 &   1.17 &   5.37 & 0.35$\pm$0.08 & 11 & 13 &   0.08$\pm$0.29 \\ 
X47 & 16:42:03.380 & 36:24:15.48 &   0.42 &   5.55 & 4.93$\pm$0.31 & 126 & 174 &   0.16$\pm$0.08
\enddata
\tablecomments{For each source we give its ID, coordinates and position error (columns 1 to 4). We report the distance to the center in column 5. We then give the net count rate, the number of counts in the 2 bands $\mathrm{B1} = 0.3-1.5$~keV and $\mathrm{B2} = 1.5-8$~keV, and the hardness ratio HR = (B2 -- B1) / (B1 + B2).}
\end{deluxetable*}

In the \chandra\ X-ray Observatory archive, we found two observations of \object{M13} with the ACIS-S detector on 2006 March 10 and 11 (MJD 53804--53805) for 27 and 28~ks with a gap of $\sim$80~ks between the observations (ObsID \dataset[ADS/Sa.CXO#obs/07290]{7290} and  \dataset[ADS/Sa.CXO#obs/05436]{5436}).

We reprocessed these datasets with CIAO 4.1 and combined the 2 datasets with the script \textit{merge\_all}. We then used \textit{wavdetect} to detect sources in the energy ranges $0.5-2$ and a broader $0.3-8$~keV band with scales of 1.0, 1.4, 2.0, 2.8, 4.0, and 5.6 pixels. We selected a threshold probability of $10^{-6}$, designed to give one false source per $10^6$ pixels. We found 47 X-ray sources in a $9'\times9'$ region centered on \object{M13} (Table~\ref{tab:chsrclist}). Two sources are only detected in the energy range $0.5-2$~keV (X14 and X30) and 12 are only detected in the broad band (including X3, X10 and X13 inside the half-mass radius). This list includes the 14 sources detected previously with \xmm\ \citep{GBW03} and available in the 2XMM catalog \citep{Watson+09}.
An image that covers the half-mass radius of \object{M13} is shown in Figure \ref{fig:xmap}. A hardness ratio diagram is given in Figure \ref{fig:xhrd}.

In the energy range $0.5-2$~keV, a total of 15 sources are located inside the half-mass radius, and 8 of them are inside the core radius. Within a circle with a radius of $4\farcm2$, we detected 35 sources ($0.5-2$~keV band). Using the study of \citet{HMS05}, which gives an estimate of the space density of background sources (active galactic nuclei, AGN, in the soft X-ray band $0.5-2$~keV), we estimated the expected number of AGN for several regions (see Table~\ref{tab:xexpect}). 
We assumed that the minimum flux of a detected source during the observation corresponds to 5 counts. A clear excess of detections is observed in the direction of M13.
We limited this study to a region of radius $4\farcm2$ where the sensitivity drops to 86\% of the central sensitivity and the PSF is slightly more elongated. As we assumed a constant minimum flux in this region, this could have led to a possible deficit of detections compared to the expected number of background sources in the larger region. However, this deficit is not observed, strengthening the excess of detections observed which may thus be slightly underestimated.
As \object{M13} is at a Galactic Latitude of $40.91^\circ$ we expect only few foreground X-ray sources. Therefore, most of the X-ray sources inside the half-mass radius are likely to be associated with \object{M13}, and a tentative estimate of $19.6\pm3.9$ sources could be associated with the cluster.

\section{Detailed results for the variable star 4}
\label{DN}

\subsection{X-ray counterpart: lightcurve and spectrum}

The star 4 falls in the error radius of an X-ray source detected with \xmm\ on 2002 January 28 and 30 \citep{GBW03}. The X-ray source is described as variable by at least a factor $\sim$2 on timescale of years by \citet{GBW03} as it should have been detected by previous X-ray observations with \rosat\ in 1994 \citep{Verbunt01}.
The 2XMM catalog \citep{Watson+09} gives a flux in the band 0.2--12.0~keV of $3.9 \pm 0.7 \times 10^{-14}$~erg~s$^{-1}$~cm$^{-2}$ ($8.8\times10^{31}$~erg~s$^{-1}$ at the distance of \object{M13}).

The position of the corresponding \chandra\ source (X6, see \S\ref{chandra}) is $\mathrm{RA}=16^h41^m42^s.48$, $\mathrm{Dec}=+36^\circ28'07\farcs6$ (J2000). We computed the 95\% error from the formula given by \citet{Hong+05} and found $0\farcs32$. Including the absolute pointing error of \chandra\ ACIS-S ($0\farcs7$ at 95\%) we find a total error on the position of $0\farcs77$ at 95\% confidence level. The variable star 4 lies $1\farcs15$ away with a positional error of $1\farcs2$ at 95\%.

\begin{figure}[tbp]
\centering
\vspace{0.5cm}
\includegraphics[width=\columnwidth]{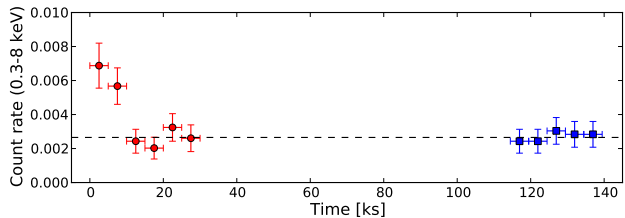}
\caption{X-ray lightcurve of \chandra\ source X6 in M13. The lightcurves start time is 2006-03-09 23:01:13 UT (MJD 53803). The mean level at low luminosity is indicated by a dashed line.}
\label{fig:xlc}
\end{figure}

\begin{figure}[tbp]
\centering
\vspace{0.5cm}
\includegraphics[height=\columnwidth,angle=-90]{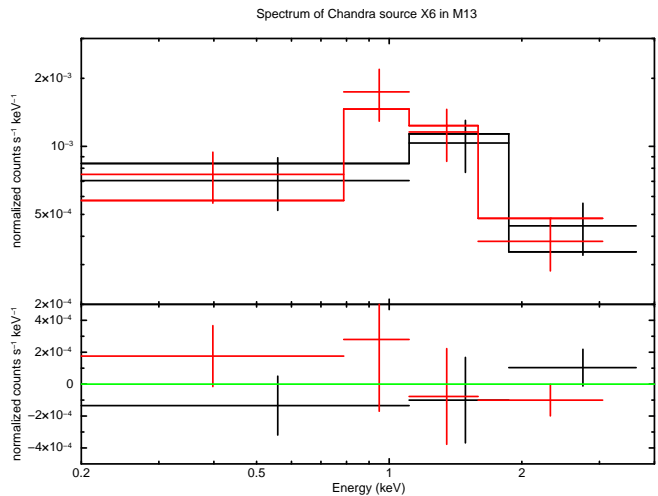}
\caption{X-ray spectrum of \chandra\ source X6 in M13. The solid lines are the absorbed bremsstrahlung model. The residuals are indicated in the bottom panel}
\label{fig:xsp}
\end{figure}

We extracted the lightcurve of source X6 from the \chandra\ data using the CIAO tool \textit{dmextract} (see Figure~\ref{fig:xlc}). The count rate of source X6 was higher by a factor $\sim$3 at the beginning of the observation and for $\sim$10~ks. The background is negligible throughout the observation.

We extracted the spectra of X6 for both observations using the tool \textit{psextract}. We generated the RMF response files separately with the tool \textit{mkacisrmf}. We used Sherpa to fit both spectra simultaneously, excluding the first 10~ks of data. Given the limited number of counts (a total of 134 counts), we tried simple models with a fixed absorption of $1.1\times10^{20}$~cm$^{-2}$ to fit the spectrum. We used the TBabs model with abundances from \citet{WAM00}. Good fits were obtained for a power law with an index of $1.3\pm0.3$ ($\chi^2$/dof = 3.9/7) or a bremsstrahlung model with temperature of k$T\sim$35~keV ($\chi^2$/dof = 4.2/7). 
The temperature is however not well constrained (k$T<50$~keV at 1$\sigma$). The spectrum and best fit are reported in Figure~\ref{fig:xsp}.

Based on the power-law models, we estimated that the flux of the source in the band 0.3--8.0~keV (with $1\sigma$ error) for the second observation is ${2.2\pm0.8\times10^{-14}}$~erg~s$^{-1}$~cm$^{-2}$ ($5.0\times10^{31}$~erg~s$^{-1}$ at the distance of \object{M13}). During the increase of counts in the first observations, the flux reached ${5.2\pm1.9\times10^{-14}}$~erg~s$^{-1}$~cm$^{-2}$ ($1.1\times10^{32}$~erg~s$^{-1}$ at the distance of \object{M13}).

Due to the low number of counts, we estimated the median energy at 3 different time intervals to look for possible spectral variability \citep[see][for error estimations]{HSG04}. During the first $\sim10$~ks, the median energy is $1.29\pm0.12$~keV, then it rises to $1.77\pm0.25$~keV at the end of the first observation, possibly indicating a slight hardening of the source in the low luminosity state. In the second observation the median energy is $1.26\pm0.16$~keV.

\subsection{Maximum U-band magnitude}

\begin{figure}[tbp]
\centering
\includegraphics[width=\columnwidth]{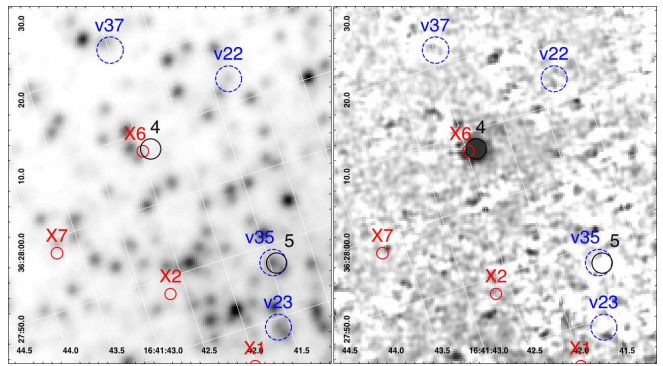}\\
\caption{Close up Faulkes U-band image around the \chandra\ source X6 and the variable star 4. Variable star candidates and X-ray sources are indicated by their 95\% error radius (black and red respectively), and known variables are indicated with a blue dashed circle of radius $1\farcs8$ for clarity. \textit{Left}: reference image. \textit{Right}: combination of subtracted images as presented in the text. We also discuss in the text complex features seen in the background.}
\label{fig:mapzoom}
\end{figure}

We averaged the subtracted images taken between MJD 53419 and 53433 with the FTN, when the star was at a maximum of luminosity. We also averaged the images between MJD 53472 and 53477 during the minimum.
In those two images, the source appears as positive and negative, respectively, compared to the emission of the source in the reference image. In order to obtain the total flux of star 4 between the minimum and maximum states, we subtracted the negative image from the positive image.
In Figure~\ref{fig:mapzoom}, we show the reference image and the obtained combination of subtracted images around star 4. An excess is clearly seen, and the error circle of the variable star and the X-ray source overlap, indicating that the two sources are counterparts. One can note that the background is irregular, with many features that show positive and negative peaks. Those are residuals from the subtraction of stars. They can be ruled out as variable stars as the integrated flux in a sufficiently large circle generally remains close to zero and well below the standard deviation of the surrounding background. The shape of the emission is also not as point-like as for star 4, and does not have the expected PSF size.

We studied the stack of subtracted images in the U-band with IRAF \textit{imexamine} in order to determine the  magnitude of the star. 
We first set the zero point with the reference image. We used 6 isolated and bright reference stars with a known spectral type from SIMBAD. These are not classified as variable and possess B and V magnitude measurements. Based on this information, we extrapolated the expected U magnitude of those stars, and scaled the measured U magnitude accordingly. The resulting standard deviation for the set of reference stars is $\sigma_{\mathrm{U}}=0.3$. As the extinction is low in this field, $\mathrm{E(B-V)}=0.02$, it is unlikely to have an effect on this measurement.

Based on the scatter of the background in one subtracted image, the size of the FWHM and the zero point, we estimate that the 5$\sigma$ detection threshold corresponds to $\mathrm{U}=19\pm0.3$.
We derived a U magnitude of $17.3\pm0.3$ for star 4. 
The star has not been reported before, so it is likely that the star was fainter and below the detection threshold during its minimum of luminosity. 
The minimum flux would then be more than $1.7$ magnitudes lower than the maximum flux. Given the large error bars, we can assume that the measured magnitude in the combination of subtracted images can be directly taken as the maximum magnitude reached by the star. 
The absolute magnitude is M$_\mathrm{U}=3.0\pm0.3$ if the object is associated with \object{M13}.

\subsection{Search for a counterpart with \hubble}


\begin{figure*}[tbp]
\centering
\includegraphics[width=\textwidth]{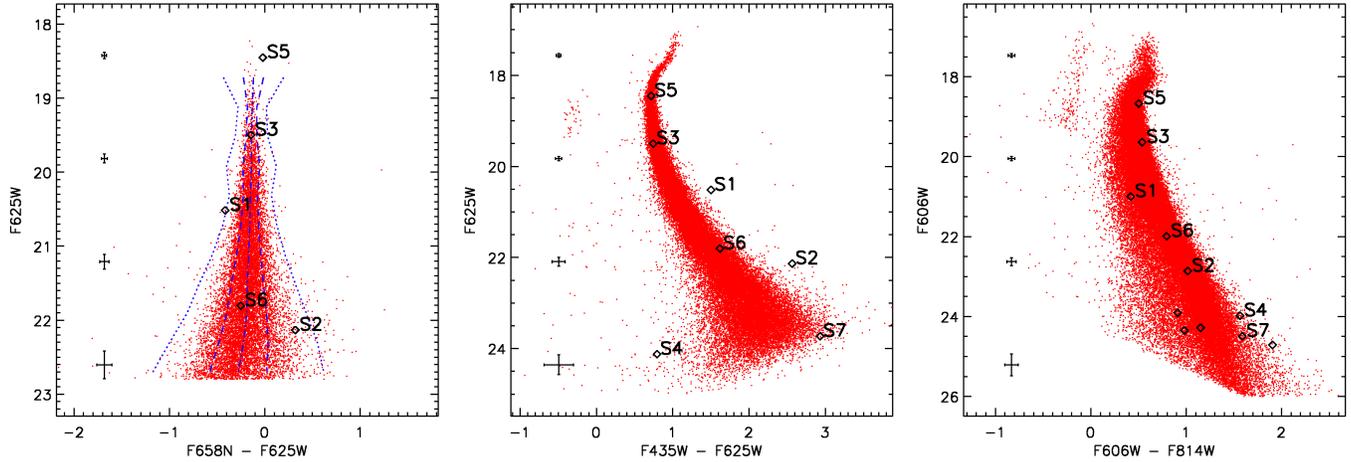}\\
\caption{Color magnitude diagrams of M13 from HST ACS/WFC images. When detected, the stars S1 to S7 shown in Figure~\ref{fig:hst} are indicated. Errors bars are shown on the left of the diagram, which correspond to the mean error for stars in the corresponding magnitude range. In the first diagram, the mean, 1$\sigma$ and 3$\sigma$ envelopes of the distribution are shown as dashed lines.}
\label{fig:hstcmd}
\end{figure*}

\begin{figure}[bp]
\centering
\includegraphics[width=\columnwidth]{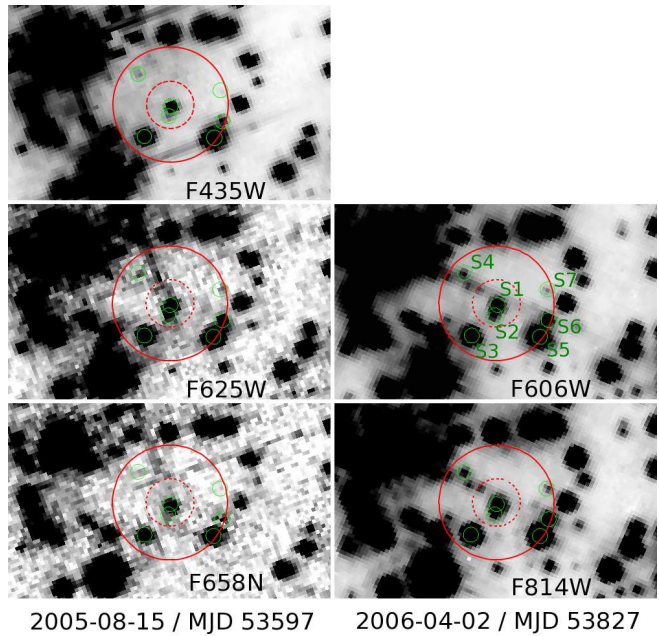}
\caption{HST ACS/WFC snapshots. Images from the datasets with proposal ID 10349 and 10775 are on the left and right columns respectively. The size of each image is $4\farcs2\times2\farcs5$, North is up, East left. The 95\% \chandra\ error on the position of the candidate CV are indicated as red circles (small dashed: only detection error, large plain: including absolute pointing error). S1 to S7 are shown with a 2 pixels radius circle.}
\label{fig:hst}
\end{figure}

\begin{deluxetable}{c@{~}ccll}
\tablecaption{\hubble\ Space Telescope archived observations.\label{tab:hstobs}}
\tablewidth{0pt}
\tablehead{\colhead{Prop. ID} & \colhead{Date} & \colhead{MJD} & \colhead{Filter} & \colhead{Exp. time}
}
\startdata
10349  & 2005-08-15    & 53597   & F435W (B)            & 2$\times$680s, 1$\times$120s \\
           &                     &            & F625W (R)            & 1$\times$360s, 1$\times$20s \\
           &                     &            & F658N (H$\alpha$) & 1$\times$800s, 1$\times$690s \\
10775  & 2006-04-02    & 53827   & F606W (R)            & 1$\times$567s \\
           &                     &            & F814W (I)              & 1$\times$567s
\enddata 
\end{deluxetable}

\object{M13} has been observed with the \hubble\ Space Telescope for 6 programs after 1995. We checked the previews of the images acquired and found 3 programs
(proposal IDs \dataset[ADS/Sa.HST#J9L957010]{10775}, \dataset[ADS/Sa.HST#J95430011]{10349}, and \dataset[ADS/Sa.HST#U5FC0109R]{8174})
where the region of the X-ray source X6 is covered. The dataset with proposal ID 8174 is composed of short exposures only (e.g. 26~s) so we focused on the 2 other datasets obtained with the Advanced Camera for Surveys (ACS) and its Wide Field Camera (WFC) detector. The list of available exposures is given in Table~\ref{tab:hstobs}.

We downloaded the pipeline-reduced images (drz files) and corrected the astrometry with GAIA using 3 reference stars in the field of view and out of the crowded region of the cluster from the UCAC2 catalog, giving an RMS of $0\farcs0003$. The error thus mainly comes from the UCAC2 reference stars, which is $0\farcs04$ (at 1$\sigma$) in this field of view. 
Images of the region around star 4 are given in Figure~\ref{fig:hst}. 

We detected the stars and performed the photometry measurement with IRAF/DAOPHOT. We used 6 isolated, non-saturated stars to create the PSF model for each image, and applied the task \textit{allstar} to fit the stars per group of at most 60 stars. The photometry was calibrated using the zero points provided on the ACS webpage\footnote{http://www.stsci.edu/hst/acs/analysis/zeropoints} to get Vega magnitudes in each filter.
We detected 11 counterparts inside the \chandra\ error circle of X6 in the F814W image, and only 5 in the F658N image. We obtained magnitudes in the 4 broad bands for 7 stars, labelled S1 to S7 according to their distance to X6, which are indicated in Figure~\ref{fig:hst}. They are also reported in the 3 different color-magnitude diagrams (CMDs) shown in Figure~\ref{fig:hstcmd}. Each diagram was generated with 2 images taken during the same program.

We note that some spikes from bright stars are apparent in the images (see Figure~\ref{fig:hst}). Unfortunately, these spikes likely contaminate S1, S2 and possibly S4 in the first dataset (F625W and F658N), and S4, possibly S1 in the second (F606W and F814W). This may explain the extreme red color of S1 and S2 in the CMD (F625W, F435W--F625W) which are then bluer and fainter relative to the main sequence in the CMD (F606W, F606W--F814W). Maybe the blue color of S4 in the CMD (F625W, F435W--F625W) is also due to contamination as the star is faint. There remains, however, a possibility that S1, S2 and S4 showed color variability.

The CMD (F625W, F658N--F625W) can be used as an indicator of H$\alpha$ excess. We note that compared to the distribution, S1 seems to present a $\sim$3$\sigma$ H$\alpha$ excess. There does not seem to be a strong contamination in the F658N band, and it is likely that it would be compensated or dominated by the contamination in the F625W band.

\section{Discussion}
\label{discussion}

\subsection{Star 4: a DN outburst}

Using observations from different observatories in X-ray and optical, we find compelling evidence that the variable star 4 (X-ray source X6) is a DN. 
First, the star shows variability on a timescale of months, consistent with DN outbursts that could last at least 24 days and reached a maximum absolute U magnitude of $\mathrm{M_U}=3.0\pm0.3$ if it is located in M13. We do not have color information, but we note that known DNe in outburst can have U--V values between 0 and --1 \citep{Harrison+04}. 
Therefore star 4 is likely to have a maximum magnitude of $\mathrm{M_V}\gtrsim3.0$.
Previously studied DNe have a maximum absolute magnitude in outburst of $\mathrm{M_V}\sim2.7-5.2$ \citep{Harrison+04}. 
The optical brightenings of the variable star 4 would thus be consistent with bright outbursts from a DN located in the core of \object{M13}.

The position of this star is consistent with an X-ray source detected using \rosat, \xmm\ and \chandra. 
The X-ray source has varied in flux between the \rosat\ and \xmm\ observations as expected for CVs, but many other X-ray sources show variability.
The emission of the \chandra\ X-ray source is compatible with the expected emission of a DN. 
Indeed, DN have luminosities in the range $10^{29-33}$~erg~s$^{-1}$ and bremsstrahlung temperatures of k$T\sim10$~keV \citep[e.g.][]{Verbunt+97,Baskill+05}. It is thus similar in luminosity to the brightest DN such as U~Gem or SU~UMa \citep[e.g.][]{Baskill+05}. We note that the temperature we estimated is also consistent with the higher temperatures found for intermediate polars, e.g. 20--30~keV \citep[e.g.][]{EI99}.
A variation in the X-ray count rate of a factor 3 has been observed with \chandra. 
This kind of fluctuation is common for CVs or magnetically active binaries. However the latter generally show an exponential decay of the X-ray flux over a time range of typically 10~ks which is not seen here.
We note that this X-ray drop occurred 2 days before U-band observations with the FTN which show the star in a bright state. 
This behavior seems consistent with the X-ray/optical anticorrelation observed for some DNe. For example, SU UMa showed a sudden drop in X-rays by a factor 4 about half a day after the optical rise \citep{CW10}.
The hard X-ray emission of SS Cyg \citep{MPT04} is almost suppressed during the optical outburst. This is somewhat contrary to the possible hardening of the source that we observed after the drop in flux. However, SS Cyg also showed an intermediate state where similar hardening and decrease of the X-ray flux occurred \citep{MPT04}. SS Cyg also showed an X-ray flux increase at the beginning of the optical outburst for about half a day, as expected from boundary-layer models \citep{WMM03}. It is possible that the drop in flux we observed corresponds in fact to this increase that precedes the optical outburst.

Finally, using archived data from the \hubble\ Space Telescope we found $\sim$11 stars inside the \chandra\ error circle. No bright star was found, thus the object is likely to be in quiescence during those observations.
One of the brightest counterparts (S1) presents a possible H$\alpha$ excess. This kind of emission could come from an accretion disk around a white dwarf and would indicate that S1 is the true counterpart. However, S1 does not show a blue excess as it is generally expected from DNe, but we note that some DNe in quiescence in GCs are consistent with main sequence stars \citep[e.g. CV2 in M22,][]{Pietrukowicz+05}. 
We note that the star S4 is suspiciously blue in one observation and could have shown color variability, however, it was close to a spike from a bright star and measurements may not be reliable.
We could measure magnitudes and colors for the 7 brightest counterparts only. Thus, one of the fainter counterparts, for which we could not determine reliable colors, could turn out to be a more convincing association. 

If we consider the most studied DNe, their absolute magnitude in quiescence lie in the range $\mathrm{M_V}=6-10$ \citep[e.g.][]{Patterson10}. The amplitude of the outburst is generally $\mathrm{d_{M_V}}=2-6$. DNe with brighter outbursts also tend to have lower amplitude. 
At the distance of \object{M13}, the quiescent magnitude range would be $\mathrm{V}=20.2-24.2$.
This outburst being bright, we would then expect a counterpart in quiescence to have a V magnitude in the brighter part of this range. This makes it less likely for fainter stars such as S4 to be the true counterpart, while S1 stands as a good candidate. However, further observations are still needed to locate the true counterpart, with better signal to noise ratio and less contamination from bright stars in order to confirm an H$\alpha$ excess.

\subsection{Population of CVs in M13}

Based on the encounter rate in M13 and the correlation of this parameter with the number of X-ray sources reported by \citet{PH06}, we expect about one source within the half-mass radius of M13 above a luminosity threshold of $4\times10^{31}$~erg~s$^{-1}$, where cluster X-ray sources are dominated by CVs. Inside the half-mass radius, we found 2 X-ray sources above this threshold. One of them is very soft and already reported as a quiescent LMXB candidate \citep{GBW03}. The other is the DN discovered in this work. This is consistent with the estimation and strengthens the identification of X6 as a CV.

\citet{Ivanova+06} performed simulations based on different CV formation channels. Their low-density cluster is the closest to \object{M13}, and they estimated that a total of 156 CVs can be formed in such a cluster. From this simulation, about 40\% of the CVs formed are purely primordial CVs (i.e. no dynamical interaction were needed to lead to their formation). Another 46\% of the CVs are also primordial, but entered the core of the cluster before the common envelope phase. Only 5\% of the CVs in the simulation come from formation channels that involve collisions or binary exchange mechanisms. Therefore, the population of CVs in \mbox{\object{M13}} could more closely resemble the population of field CVs than populations in denser GCs, where dynamical effects are believed to modify and enhance the production of CVs.

There have been few detections of DNe in GCs, with only 13 DNe reported so far in the 157 Galactic GCs \citep[e.g.][]{Pietrukowicz+08,KT09}. 
Based on our knowledge of field CVs, which is extremely biased, it is tempting to try to estimate the expected number of DN outbursts we should have detected in M13, for example following \citet{Shara+96}.
The DN population constitutes about half of the known CVs listed in the catalog of \citet{Downes+01}, which contain mostly field CVs. We could then expect $\sim$78 DNe in M13, based on those rough estimations.
The mean duty cycle, or probability of a DN being in outburst at a random epoch, has been found to be $\sim$15\% based on the observation of 21 DNe \citep{SM84}. We can group our observations into 5 epochs of few days, thus the likelihood of a DN being in quiescence in those 5 epochs is $0.85^{5}=0.44$. Thus $\sim$42 DNe should have been in outburst in at least one of the epochs. 
It is then harder to estimate the number of outbursts that would have been bright enough to be detected in our observation. Our detection limit at $\mathrm{U}\sim19$ would convert to $V\sim20$ ($\mathrm{M_{V}}\sim5.7$) due to the general blue color of DN outbursts \citep{Harrison+04}. We can then use the distribution of V magnitudes at maximum light from \citet[][]{Patterson10} based on the observations of $\sim$45 DNe, which shows that we could have expected to detect at least two third of those DNe in outburst at the distance of M13. Finally, $\sim28$ DNe in outburst would have been expected when we detect only one.

However, it is important to note that this estimate is based on several approximations, which makes this number likely to be well over-estimated.
If we suppose that the mean duty cycle of DNe is 1\% instead of 15\%, we then expect $\sim$2 DNe in outburst. We can also suppose that the proportion of DNe is much less than half of all CVs. It would have to be less than 5\% to expect so few DNe in eruption, as in our observations. Considering that M13 is not an extremely dense GC, with most of its CVs likely being primordial, we may be able to deduce basic information on the composition of the global Galactic CV population. Even if there are large errors in each of our estimates, it seems likely that the mean duty cycle of DNe is lower than 15\%, and/or the intrinsic fraction of DNe among CVs may be smaller than catalogs suggest. 
If the predictions of \citet{Ivanova+06} regarding total numbers of CVs in GCs are correct, then a possible solution to the apparent discrepancy between M13 and the field could be that there is a sub-population of CVs that undergo outbursts but have lower luminosities and longer outburst recurrence times than those of the currently known field DNe.
       
Several population synthesis models already showed that a sub-population of faint CVs with long outburst recurrence times may dominate the total number of CVs \citep{kolb93,howell+97,willems05}. However there are still few examples of such a sub-population \citep[e.g.][]{gaensicke09}.
Deeper surveys of the Galactic Plane are thus needed to catch fainter CVs in order to test this hypothesis. Having observations on timescales of decades could also help to detect outbursts with longer recurrence times. This will maybe be possible through the DASCH project which aims to digitize half a million photographic plates obtained over 100 years in the last century \citep{grindlay09,laycock10,Servillat+11}.

On the other hand, it is possible that our observations were inefficient in detecting DN outbursts.
Considering the expected recurrence time of months to years and a duration of at maximum a few weeks for DN outbursts, the optical observation coverage that is likely to detect most of the DN outbursts requires both short scale observations (over few nights) and a recurrence every few weeks, if possible over a year. This kind of survey is thus time consuming, particularly for the large telescopes that are needed if the target is crowded as it is the case for GCs. 

Future variability surveys, such as the VISTA Variables in The Via Lactea \citep[VVV,][]{Minniti+10} public survey and the Large Synoptic Survey Telescope \citep[LSST,][]{Ivezic+08} will conduct an even denser coverage with a 4m and 8.4m telescopes, respectively. One objective would be to uncover more DNe in GCs and in the field, in order to base our study of CV populations on samples with a decent number of objects.


\section{Summary and Conclusions}

We performed a search for optical outbursts in the GC M13. We found one star in M13 which showed variability consistent with outbursts. This star is consistent with being a DN, based on the optical variability and flux, and on the X-ray variability, spectrum and luminosity. Several possible counterparts were detected in \hubble\ images, including one with a possible H$\alpha$ excess and consistent magnitude in quiescence.

The fact that we detect one DN when more could have been expected suggests that the known population of local CVs is biased, and short-period, faint, low-duty-cycle (1\%) DNe are likely to dominate the global CV population, as expected from population synthesis models. Their proportion is however not well constrained and future variability surveys will help to better characterize the global CV population.


\acknowledgments

We thank the anonymous referee whose careful review helped to clarify this paper.
MS acknowledges supports from NASA/\chandra\ grants AR9-0013X, GO0-11063X and NSF grant AST-0909073.
This research has made use of the SIMBAD database, operated at CDS, Strasbourg, France.\\



{\it Facilities:} \facility{FTN}, \facility{CXO (ACIS)}, \facility{HST (ACS)}.

\end{document}